\documentclass[amssymb,prb,aps,twocolumn,amsmath,showpacs]{revtex4}

\usepackage[dvips]{graphicx}
\usepackage{amstext}

\begin{document}

\title{Area-dependence of spin-triplet supercurrent in ferromagnetic Josephson junctions}
\author{Yixing Wang, W. P. Pratt, Jr., and Norman O. Birge}
\email{birge@pa.msu.edu} \affiliation{Department of Physics and
Astronomy, Michigan State University, East Lansing, Michigan
48824-2320, USA}

\date{\today}

\begin{abstract}
Josephson junctions containing multiple ferromagnetic layers can
carry spin-triplet supercurrent under certain conditions.
Large-area junctions containing multiple domains are expected to
have a random distribution of 0 or $\pi$ coupling across the
junction surface, whereas magnetized samples should have uniquely
$\pi$ coupling everywhere. We have measured the area dependence of
the critical current in such junctions, and confirm that the
critical current scales linearly with area in magnetized
junctions.  For as-grown (multi-domain) samples, the results are
mixed.  Samples grown on a thick Nb base exhibit critical currents
that scale sub-linearly with area, while samples grown on a
smoother Nb/Al multilayer base exhibit critical currents that
scale linearly with area.  The latter results are consistent with
a theoretical picture due to Zyuzin and Spivak that predicts that
the as-grown samples should have global $\pi/2$ coupling.
\end{abstract}

 \maketitle

\section{Introduction}

When a superconducting (S) metal is placed in contact with a
nonsuperconducting (N=normal) metal, Cooper pairs can "leak" out
of the superconductor and modify the properties of both materials.
This process, called the superconducting proximity effect, has
been studied for several decades.\cite{deGennes}  When the normal
metal is replaced by a ferromagnetic (F) metal, the two electrons
of the pair enter different spin bands with different Fermi
wavevectors. As a result, the pair correlations oscillate and
decay rapidly with increasing distance from the S/F
interface.\cite{Demler} The oscillating, short-ranged proximity
effect in S/F systems was predicted as early as
1982,\cite{Buzdin:82} and has been observed convincing by many
groups over the past decade.\cite{BuzdinReview}

In 2001, it was predicted that pair correlations with spin-triplet
symmetry can be generated in S/F systems containing certain kinds
of magnetic inhomogeneities involving non-collinear
magnetizations, even if all of the superconducting materials in
the system have conventional spin-singlet
symmetry.\cite{Bergeret:01,Kadigrobov:01,Eschrig:03,Bergeret:05}
Spin-triplet pair correlations do not experience the exchange
field in F, hence the proximity effect due to those pair
correlations is long ranged. Some experimental evidence for
spin-triplet correlations appeared in
2006,\cite{Keizer:06,Sosnin:06} then more convincing evidence
appeared in
2010.\cite{Khaire:10,Robinson:10,Sprungmann:10,Anwar:10}  Our own
contribution \cite{Khaire:10} was based on measurements of the
critical current $I_c$ in Josephson junctions of the form
S/F'/SAF/F''/S, where SAF stands for ``synthetic antiferromagnet"
and F' and F'' are thin ferromagnetic layers whose magnetizations
must be at least partly non-collinear with that of the SAF.  The
SAF is a Co/Ru/Co trilayer with Ru thickness 0.6 nm, which causes
anti-parallel coupling of the two surrounding Co magnetizations.
Because of that anti-parallel coupling, the SAF produces nearly
zero net magnetic flux, which would otherwise distort the
``Fraunhofer patterns" one observes in plots of $I_c$ vs magnetic
field $H$ applied in the plane of the Josephson junction.  We
found that $I_c$ in our samples hardly decreased as the total Co
thickness was increased up to 30 nm, whereas $I_c$ decayed rapidly
in similar samples without the F' and F''
layers.\cite{Khasawneh:09} The long-range nature of the
supercurrent in the samples containing F' and F'' layers provided
strong evidence for the spin-triplet symmetry of the
current-carrying electron pairs. Furthermore, we have recently
shown that $I_c$ increases further when we magnetize our samples
with an in-plane applied magnetic field.\cite{Klose:12} The
explanation is that the F' and F'' layers are magnetized parallel
to the field, while the SAF undergoes a ``spin-flop" transition
whereby the two Co layers end up with their magnetization
perpendicular to the direction of the applied
field.\cite{Zhu:98,Tong:00} According to
theory,\cite{Bergeret:03,Houzet:07,VolkovEfetov:10,Trifunovic:10}
this configuration with perpendicular magnetizations maximizes the
magnitude of the spin-triplet supercurrent.

\begin{figure}[tbh]
\begin{center}
\includegraphics[width=2.4in]{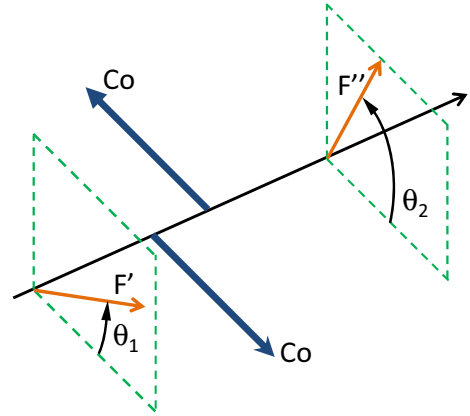}
\end{center}
\caption{Cartoon showing relative orientations of magnetization
for the ferromagnetic layers in our Josephson junctions. If angles
$\theta_1$ and $\theta_2$ have the same sign (where we constrain
$|\theta_1|,|\theta_2|<\pi$), the junction will have $\pi$
coupling; if they have opposite signs, the junction will have 0
coupling.}\label{Cartoon}
\end{figure}

The results described above raise several questions, only some of
which have been answered by subsequent work in our
group.\cite{Khasawneh:11,Klose:12}  The question that motivated
this paper arises from the theoretical prediction that Josephson
junctions of the form S/F'/F/F''/S, carrying spin-triplet
supercurrent, can be either in the 0-state or the $\pi$-state
depending on the relative orientations of the three ferromagnetic
layers.\cite{Bergeret:03,Houzet:07} (It does not matter whether
the central F layer is a single ferromagnetic layer or an
SAF.\cite{VolkovEfetov:10,Trifunovic:10}) The situation is
illustrated in Figure \ref{Cartoon}, which shows the relative
orientations of the magnetizations of all four ferromagnetic
layers in our junctions.  According to theory, if the two angles
$\theta_1$ and $\theta_2$ have the same sign, then the junction
will have $\pi$ coupling; if they have opposite signs, the
junction will have 0
coupling.\cite{Houzet:07,VolkovEfetov:10,Trifunovic:10} (We define
the angles by the constraint $|\theta_1|,|\theta_2|<\pi$.) Since
the magnetic layers in our samples consist of many domains when
the samples are first grown, we would expect the Josephson
coupling in our junctions to exhibit a random spatially-varying
pattern of 0-coupling and $\pi$-coupling across the junction area.
In that case, if a fixed difference in gauge-invariant phase is
applied across the junction, some areas of the junction will
provide positive supercurrent while others will provide negative
supercurrent -- i.e. supercurrent flowing in the opposite
direction.  One could then calculate the total supercurrent
naively using an analogy to the random walk problem: while the
mean supercurrent averaged over many domains would be zero, the
typical supercurrent in a given sample would be proportional to
the square-root of the number of domains, hence to the square-root
of the junction area. After the samples are magnetized, the
magnetizations of the F' and F'' layers are parallel to each
other, hence $\theta_1$ and $\theta_2$ have the same sign and the
junction should have $\pi$-coupling everywhere. In that case, the
critical supercurrent will be proportional to the junction area,
as is the case in conventional Josephson junctions.

A completely different view of a Josephson junction containing a
random spatially-varying pattern of 0 and $\pi$ couplings has been
proposed by Zyuzin and Spivak (ZS).\cite{Zyuzin:00} Those authors
addressed S/F/S junctions with spin-singlet rather than
spin-triplet supercurrent, and considered the situation where the
F-layer thickness is large, so that the average supercurrent is
small, whereas mesoscopic fluctuations of the Josephson coupling
have random sign.  In our samples, the spin-singlet supercurrent
is negligibly small (c.f. Figure 3 in Ref.
[\onlinecite{Khaire:10}]), and the random-sign spin-triplet
Josephson coupling arises from the local variations in magnetic
domain structure. In spite of the different mechanisms underlying
the spatially-varying random-sign Josephson coupling, there is no
apparent reason why the ZS model should not apply to our
spin-triplet Josephson junctions. ZS calculated the total energy
of such a junction, and concluded that the ground state
corresponds to, on average, a $\pi/2$ phase difference between the
two superconducting electrodes.  The phase difference is spatially
modulated, with local variations toward lower phase difference in
regions of 0-coupling and larger phase difference in regions of
$\pi$-coupling.  According to the ZS result, the total
supercurrent scales with the junction area, as is the case for
conventional Josephson junctions.

The purpose of this paper is to measure experimentally the area
dependence of the supercurrent in our Josephson junctions, to
determine which of the pictures presented above applies.

\section{Sample Fabrication}

\begin{figure}[tbh]
\begin{center}
\includegraphics[width=3.2in]{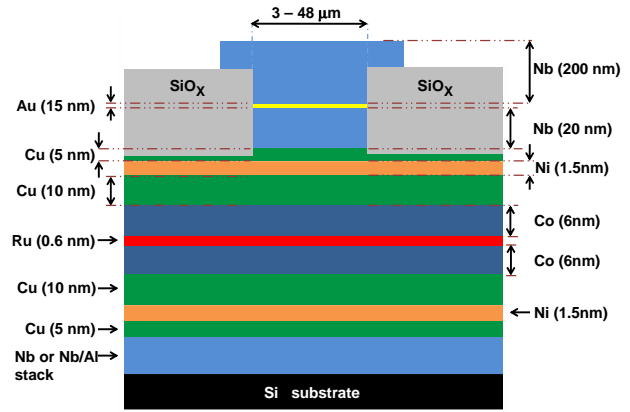}
\end{center}
\caption{(color online) Schematic diagram of Josephson junction
samples (not to scale).  The current flows in the vertical
direction. }\label{Schematic}
\end{figure}

The Josephson junctions in this work were fabricated by dc triode
sputtering, photolithography and ion-milling, as described in our
previous publications.\cite{Khaire:10,Khasawneh:09} The structure
of the junctions is shown schematically in Figure \ref{Schematic}.
In this work, we have grown two types of samples with different
superconducting base layers: either a single 150-nm layer of Nb,
or a Nb/Al multilayer stack described below.  For the first kind
of sample we start by growing a multilayer of the form
Nb(150)/Cu(5)/Ni(1.5)/Cu(10/Co(6)/Ru(0.6)/Co(6)/
Cu(10)/Ni(1.5)/Cu(5)/Nb(20)/Au(15),
where all thicknesses are in nm.  That stack is sputtered in one
run without breaking vacuum.  For the second type of sample, the
Nb base layer was replaced with a Nb/Al multilayer of the form
[Nb(40nm)/Al(2.4nm)]$_3$/Nb(40nm)/Au(15nm).  This structure was
motivated by the work of Thomas et al.\cite{Ketterson:98}, who
used a thin Nb/Al multilayer on top of thick Nb to reduce surface
roughness.  For those samples, the chamber was opened briefly
after the Au deposition, while flowing N$_2$ gas, to change
sputtering targets. (Our sputtering system holds 6 targets,
whereas the second type of samples require 7 different materials.)
After pumpdown, we sputtered first 20 nm of Nb, followed by the
rest of the stack up through the top Au layer.  We know from our
top electrode fabrication procedure that the top 15-nm Au layer
adequately protects the underlying Nb from oxidation, and is
driven superconducting when sandwiched between two Nb layers.  The
same should be true for the bottom Au layer protecting the Nb/Al
multilayer base.

\begin{figure}[tbh!]
\begin{center}
\includegraphics[width=2.8in]{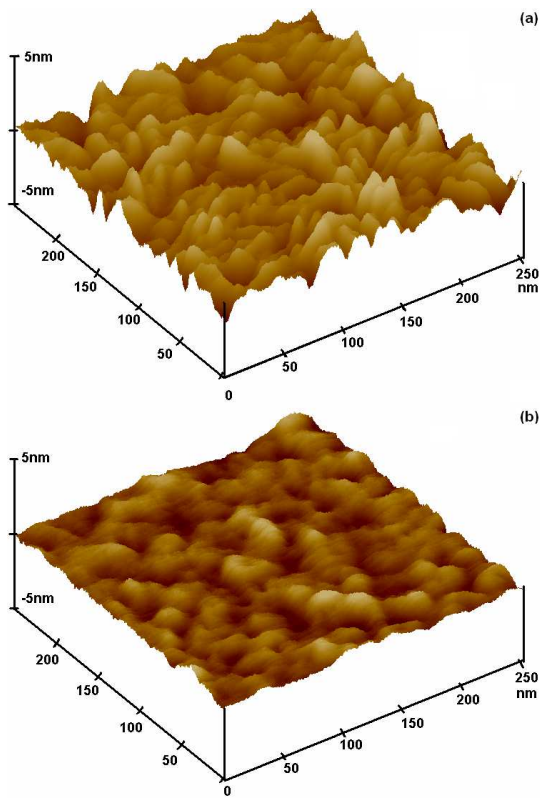}
\end{center}
\caption{(color online) Atomic force microscopy pictures of (a) a
200-nm thick Nb base layer and (b) a Nb/Al multilayer as described
in the text.} \label{AFM_results}
\end{figure}

To ascertain the surface roughness of the two types of base
layers, we performed atomic force microscopy (AFM) measurements on
a bare 200-nm Nb base and on a Nb/Al multilayer stack (up to the
Au layer discussed above).  The results are shown in Figure
\ref{AFM_results}. The root-mean-squared roughnesses of the first
and second base layers are 0.53 nm and 0.23 nm, respectively, over
the $250 \times 250$nm$^2$ area shown.  As expected, the Nb/Al
multilayer provides a smoother base than the pure Nb.

For both types of samples, Josephson junctions with circular cross
section were defined by photolithography and ion milling. Because
we wanted to obtain data on multi-domain samples covering a large
dynamic range of areas, we fabricated junctions with diameters of
3, 6, 12, 24, and 48 micrometers. Unfortunately, the largest
samples rarely produced high-quality data, hence we restrict
ourselves here to the samples of diameters 3, 6, and 12 $\mu$m.
Another difference between the samples measured here and those
measured in our previous publications is that these were ion
milled only to the top copper layer in order to keep the magnetic
domain structure intact, whereas our previous samples were
typically ion milled partway through the top Co layer.

\section{Experimental Results}

All of the data reported here were acquired at 4.2K with the
sample dipped into a liquid helium dewar.  A current comparitor
circuit using a Superconducting Quantum Interference Device
(SQUID) as a null detector was used to measure the current-voltage
(I-V) characteristic of the samples.\cite{EdmundsPratt}  All
samples exhibit the standard I-V characteristic for an overdamped
Josephson junction:
\begin{equation}\label{I_vs_V}
V(I) = Sign[I]*R_N \textrm{Re}[(I^2-I_c^2)^{1/2}]
\end{equation}
where $R_N$ is the normal-state resistance determined from the
slope of the V-I relation at large currents.

\subsection{Fraunhofer patterns and result of magnetizing samples}

\begin{figure}[tbh]
\begin{center}
\includegraphics[width=3.2in]{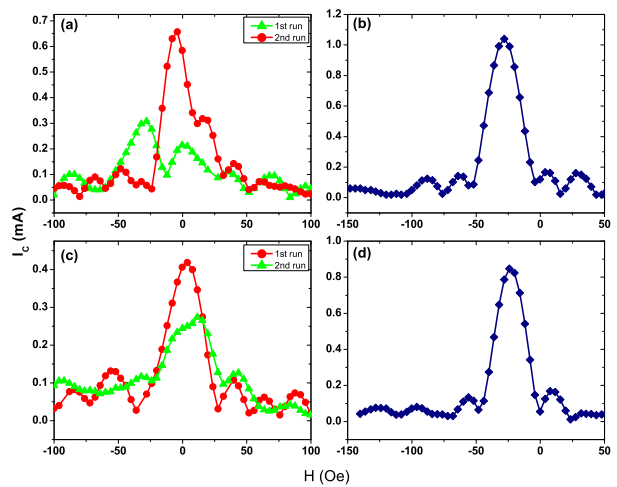}
\end{center}
\caption{(color online) Critical current vs. applied field for two
3-$\mu$m diameter Josephson junctions, measured in the virgin
state (panels a and c), and after the samples were magnetized by a
large in-plane field (panels b and d).  In the virgin state, two
separate runs are shown for each sample.  The lines connect the
data points; they are only guides for the eye.}\label{Fraunhofers}
\end{figure}

All samples are initially characterized by applying a small
magnetic field perpendicular to the current direction, i.e. in the
plane of the substrate.  A plot of $I_c$ vs $H$ should yield the
classic ``Fraunhofer pattern" (actually an Airy pattern for our
circular pillars).  Figure \ref{Fraunhofers} shows representative
$I_c$ vs $H$ data for two 3-$\mu$m-diameter junctions, in the
virgin state (panels a and c), and after being magnetized by a
large in-plane field (panels b and d).  We measured a few samples
several times to determine how much the data vary from run to run;
panels a and c show two virgin-state runs for two of these
samples.  Several features are evident from the data: i) the
Frauhofer patterns in the virgin state fluctuate from run to run;
ii) the quality of the Fraunhofer patterns is better after the
samples are magnetized than in the virgin state; iii) $I_c$ is
enhanced after the samples are magnetized, as we reported
recently;\cite{Klose:12} and iv) the central peak in the
Fraunhofer patterns of the magnetized samples is shifted to
negative field by about 30 Oe.

The variability and relatively low quality of the Frauhofer
patterns in the virgin state are undoubtedly due to the random
domain structures of the ferromagnetic layers in the
samples.\cite{Bourgeois:01,Khaire:09} One cannot know \textit{a
priori} if the problem is due to the 1.5-nm thick Ni F' and F''
layers or to the two 6-nm thick Co layers making up the SAF. We
believe it is the former, given the improvement in the quality and
reproducibility of the Fraunhofer patterns after magnetization.
Magnetizing the samples forces the Ni domain magnetizations to
point in nearly the same direction, and probably causes the
average domain size to grow. In contrast, the Co/Ru/Co SAF is
expected to have its best antiparallel coupling in the virgin
state.

The enhancement of $I_c$ by magnetizing the samples has potential
contributions from two factors.  In our recent
work,\cite{Klose:12} we emphasized optimization of the angles
$\theta_1$ and $\theta_2$ between the inner and outer
ferromagnetic layer magnetizations. Those angles vary randomly
across the junction area in the virgin-state samples, whereas they
should both be close to the optimal value of $\pi$/2 after the
samples are magnetized, due to the Co/Ru/Co SAF undergoing a
spin-flop transition. A second contributor may be the fact that,
in the virgin state, the Josephson coupling varies randomly
between 0-coupling and $\pi$-coupling across the junction area. If
the random-walk picture discussed earlier is valid, then one would
expect the value of $I_c$ in a typical sample to scale with the
square-root of the junction area, as discussed earlier. After the
samples are magnetized, there should be $\pi$-coupling everywhere,
so that the supercurrent adds constructively across the entire
junction area.

Finally, the shift in the Fraunhofer patterns after magnetization
has been observed
previously.\cite{Ryazanov:99,Khaire:09,Supplement:12} The central
peak occurs at the point where the flux due to the applied field
exactly cancels the intrinsic flux due to the magnetization of the
Ni layers inside the junction.

\begin{figure}[tbh!]
\begin{center}
\includegraphics[width=3.2in]{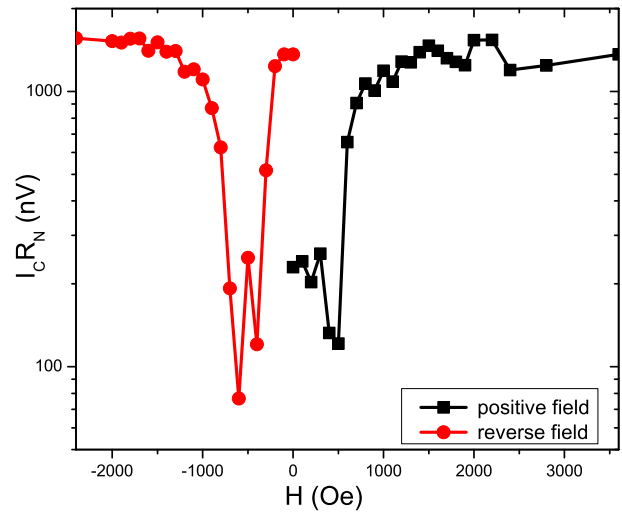}
\end{center}
\caption{(color online) $I_c R_N$ product vs. applied in-plane
field for a 6-$\mu$m diameter Josephson junction.  The sample was
first magnetized in positive field (squares), then the sample was
demagnetized and finally re-magnetized in negative field
(circles).} \label{Ic_enhancement}
\end{figure}

The evolution of $I_c$ as the sample is magnetized is shown for a
6-$\mu$m diameter Josephson junction in Figure
\ref{Ic_enhancement}.  The sample was first measured in the
as-grown state (H = 0).  Then the magnetizing field $H$ was
stepped up to 3600 Oe with varying step sizes evident in the
figure.  After application of each value of $H$, the field is
reduced to zero and the Fraunhofer pattern is measured in low
field.  The squares show the resulting values of $I_cR_N$ as the
sample is magnetized.  For low fields, nothing happens. Then there
is a shallow dip in $I_cR_N$ for $H$ near 500 Oe.  That dip is
observed in many samples, but is not fully understood; it may be
related to a change of the Ni domain structure. When $H$ is
increased above 500 Oe, $I_c$ increases sharply.  The field range
where $I_c$ increases corresponds to the field range where the Ni
films become magnetized.  (We know this because the field where
$I_c$ increases varies with Ni layer thickness, and matches the
coercive field of the Ni determined from separate magnetization
measurements of large-area Ni films.\cite{Klose:12}) Figure
\ref{Ic_enhancement} also shows what happens when a field is
applied in the opposite direction to the original magnetizing
field (circles). Again, nothing happens for small field values.
Then, as the Ni films are demagnetized, $I_c$ drops to values as
low as or even lower than the value at the dip we observed when
first magnetizing the samples. As the Ni films are re-magnetized
in the negative direction, $I_c$ increases sharply again to a
value essentially identical with that observed on the positive
field side.  We have measured full magnetization curves for
several samples, and they all look very similar to the one shown
in Figure \ref{Ic_enhancement}.

\subsection{Area dependence: samples with Nb base layer}

\begin{figure}[tbh!]
\begin{center}
\includegraphics[width=2.8in]{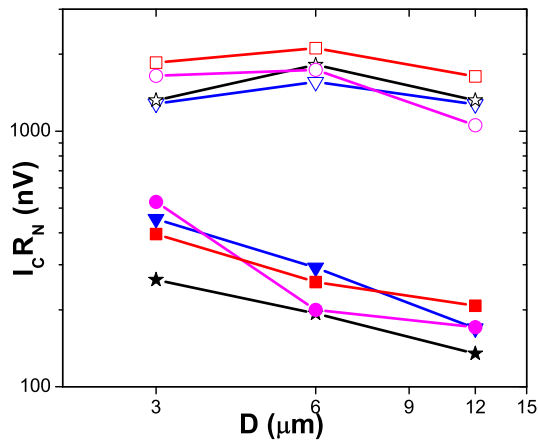}
\end{center}
\caption{(color online) Critical current times normal-state
resistance vs. junction diameter for Josephson junctions grown on
a 150-nm Nb base electrode.  Solid symbols represent virgin-state
data; open symbols represent data acquired after the samples were
magnetized by a large ($\approx$2000 Oe) in-plane magnetic field.}
\label{area_dependence1}
\end{figure}

To provide good statistics and to reveal the extent of
sample-to-sample fluctuations, we have fabricated and measured a
large number of Josephson junctions with diameters of 3, 6, and 12
$\mu$m.  Figure \ref{area_dependence1} shows the results for
samples grown on our traditional thick (150 nm) superconducting Nb
base, for both the virgin and magnetized states.  The points
representing the magnetized state (open symbols) are averages of
the measurements taken after application of 1600, 2000, and 2400
Oe.  (There is little variation of $I_cR_N$ between those three
measurements.) The points representing the virgin state (solid
symbols) are usually averaged over two runs, although  a few
samples were measured only once, and one sample was measured 5
times in the virgin state.

The results of the magnetized state measurements in Figure
\ref{area_dependence1} clearly show that $I_cR_N$ is essentially
independent of sample area. Since $R_N$ is inversely proportional
to junction area, this means that $I_c$ is proportional to area.
That is the usual situation, and is what one expects when the
Josephson coupling is uniform across the junction area.  In
contrast, the virgin-state data show a decrease in $I_cR_N$ with
increasing sample size. According to the random walk model
discussed in the introduction, $I_c$ should scale with the
square-root of the junction area, hence $I_cR_N$ should scale
inversely with the square-root of area, or equivalently, inversely
with junction diameter $D$.  The virgin-state data shown in Figure
\ref{area_dependence1} do exhibit a noticeable decrease with
junction diameter, supporting the random walk picture, although
the dependence is slightly less steep than $I_c \propto D^{-1}$.

\subsection{Samples with Nb/Al base layer}

\begin{figure}[tbh!]
\begin{center}
\includegraphics[width=2.8in]{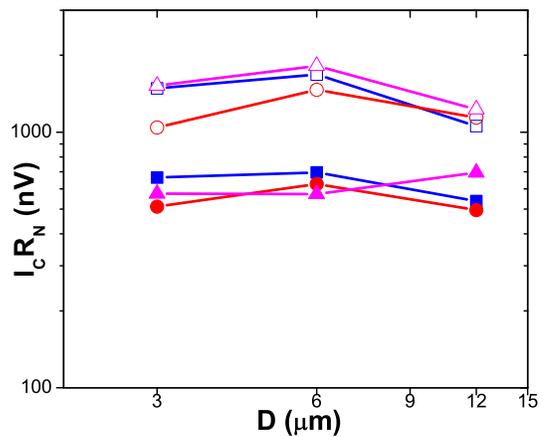}
\end{center}
\caption{(color online) Critical current times normal-state
resistance vs. junction diameter for Josephson junctions grown on
a Nb/Al multilayer as described in the text. Solid symbols
represent virgin-state data; open symbols represent data acquired
after the samples were magnetized by a large (~2000 Oe) in-plane
magnetic field.} \label{area_dependence2}
\end{figure}

As shown in Figure \ref{AFM_results}, the Nb/Al multilayer base
provides a smoother surface than the pure Nb base layer. We were
curious as to whether the smoother base would influence any of the
Josephson junctions properties.  We performed the same
measurements of $I_c$ on the second batch of samples, grown on the
smoother Nb/Al base, as were performed on the first batch, grown
on pure Nb.  The results are shown in Figure
\ref{area_dependence2}. In the magnetized state, the results agree
closely with those of the first batch, shown in Figure
\ref{area_dependence1}.  Not only is $I_cR_N$ essentially
independent of junction diameter, but the actual values of
$I_cR_N$ are very close to those of the previous batch of samples.
After being magnetized, there is remarkable consistently in the
values of $I_cR_N$ for the 21 samples displayed in Figures
\ref{area_dependence1} and \ref{area_dependence2}.  In the virgin
state, however, the story is different. In contrast to what we
observed in the first batch of samples, $I_cR_N$ in the second
batch hardly varies with junction diameter. These samples
deposited on top of the smoother base electrode appear to validate
the ZS theory, which predicts that $I_c$ should scale linearly
with junction area.

\begin{figure}[tbh!]
\begin{center}
\includegraphics[width=2.8in]{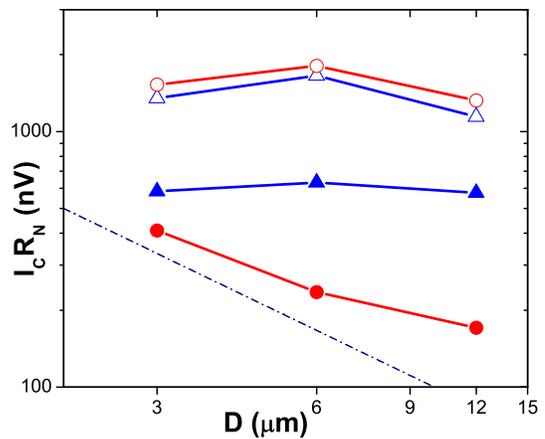}
\end{center}
\caption{(color online) Summary of $I_cR_N$ data for all the
Josephson junctions studied in this work.  Each symbol represents
the average value for all samples of a given size and base layer,
in either the virgin-state (solid symbols) or after being
magnetized (open symbols).  The circles represent samples grown on
a 150-nm Nb base layer, while the triangles represent samples
grown on a Nb/Al multilayer described in the text.  The dot-dashed
line illustrates the relation $I_cR_N \propto D^{-1}$.}
\label{average_results}
\end{figure}

The situation is summarized in Figure \ref{average_results}, where
we have averaged together the values of $I_cR_N$ for all samples
of a given diameter, fabricated on a given base layer.  With some
of the sample-to-sample fluctuations averaged out, the trends are
striking: i) The magnetized data are remarkably consistent, and
hardly depend on the base layer; ii) the virgin-state data from
samples deposited on the thick Nb base  exhibit $I_cR_N$ values
that decrease substantially with junction diameter, but not quite
as fast as $D^{-1}$, which is shown by the dot-dashed line in the
figure; iii) the virgin-state data from samples deposited on the
smoother Nb/Al multilayer base exhibit $I_cR_N$ values that are
independent of junction diameter.

What are we to make of the results shown in Figure
\ref{average_results}?  It appears that the roughness of the base
layer has a profound effect on the area scaling of $I_c$ in the
virgin state.  One can imagine several possible explanations. One
is that the spectrum of Andreev bound states, and hence the
Josephson current, in an S/F/S junction depends in a fundamental
way on whether the S/F interface is smooth or rough.  There has
been some theoretical work on interface roughness in S/N systems,
\cite{Zareyan:01} but we are not aware of any such work that
addresses our results directly.  A second possibility is that the
roughness of the Nb base layer perturbs the domain structure of
the Ni F' and F'' layers -- possibly even to the extent that one
or both of those layers are not continuous. If this is the case,
one still has to explain how modified or discontinuous F' and F''
layers would affect the area scaling of the junction critical
current. We know that the coercive fields of the Ni layers
increase continuously with decreasing Ni thickness down to 1 nm,
as shown in the data in Figure 2 of Ref. [\onlinecite{Klose:12}],
so there does not appear to be any anomalous magnetic behavior
induced by the rough base. A third possibility is that the
observed sub-linear scaling of the supercurrent with junction area
for the junctions grown on the rougher Nb base is simply a
reflection of the gradual deterioration of the quality of the
Fraunhofer patterns with increasing sample size. A possible way to
ameliorate that issue in the future would be to use PdNi alloy
rather than pure Ni as the F' and F'' layers.  PdNi is a weak
ferromagnetic material with small magnetization, and in earlier
work we were able to produce Josephson junctions with high-quality
Fraunhofer patterns even with much thicker PdNi layers than one
would need for this experiment.\cite{Khaire:09}  The optimal PdNi
thickness for producing spin-triplet supercurrent is in the range
of 4-6 nm, which is much thicker than the 1-2 nm optimal range for
Ni.\cite{Khaire:10,Khasawneh:11} We chose pure Ni for the F' and
F'' layers in the present work because it produces the largest
values of $I_c$ in the virgin state,\cite{Khasawneh:11} but
thicker PdNi layers might be less sensitive to the nm-scale
roughness of the Nb base.

\section{Conclusions and Outlook}

In summary, we have measured the area-dependence of the critical
current in S/F/S Josephson junctions carrying spin-triplet
supercurrent.  After the samples are magnetized, the critical
current has its largest value, and it scales linearly with area as
is the case for conventional Josephson junctions.  In the virgin
state, however, the results are mixed.  Samples grown on our
traditional thick Nb base exhibit critical currents that grow
sub-linearly with area, whereas samples grown on a smoother Nb/Al
multilayer base exhibit critical currents with conventional area
scaling.  The former may be an indication that the supercurrent is
not uniform over the junction area, while the latter provides
indirect support for a theoretical model of Zyuzin and
Spivak.\cite{Zyuzin:00}

In the future, it would be interesting to measure the
current-phase relation of our junctions, both in the virgin and
magnetized states.  According to the ZS theory,\cite{Zyuzin:00}
the virgin state junctions should be in a $\pi/2$ state, while
according to spin-triplet junction
theory,\cite{Houzet:07,VolkovEfetov:10,Trifunovic:10} the
magnetized junctions should be in the $\pi$ state.  Current-phase
measurements are technically more challenging than the
critical current measurements reported here, but
they might provide more direct evidence of the underlying physics
than do the area-dependent measurements reported area.

Acknowledgments:  We acknowledge helpful conversations with H.
Hurdequint, K. Michaeli, and B. Spivak. We also thank R. Loloee
and B. Bi for technical assistance, and use of the W.M. Keck
Microfabrication Facility. This work was supported by the U.S.
Department of Energy under grant DE-FG02-06ER46341.

\end{document}